\documentclass[conference]{IEEEtran}
\IEEEoverridecommandlockouts
\usepackage{cite}
\usepackage{amsmath,amssymb,amsfonts}
\usepackage{algorithmic}
\usepackage{graphicx}
\usepackage{textcomp}
\usepackage{xcolor}
\def\BibTeX{{\rm B\kern-.05em{\sc i\kern-.025em b}\kern-.08em
    T\kern-.1667em\lower.7ex\hbox{E}\kern-.125emX}}

\def\BibTeX{{\rm B\kern-.05em{\sc i\kern-.025em b}\kern-.08em T\kern-.1667em\lower.7ex\hbox{E}\kern-.125emX}}

\begin{document}

\title{Analyzing Ultra-Low Inter-Core Crosstalk Fibers in Band and Space Division Multiplexing EONs (Pre-Print)\\
\thanks{Farhad Arpanaei acknowledges support from the CONEX-Plus programme funded by Universidad Carlos III de Madrid and the European Union's Horizon 2020 research and innovation programme under the Marie Sklodowska-Curie grant agreement No. 801538. The authors from UC3M would
like to acknowledge the support of the EU-funded SEASON project (grant No.101096120) and Spanish-funded Fun4date-Redes project (grant No.PID2022-136684OB-C21).}}

\author{\IEEEauthorblockN{F. Arpanaei$^{1,2}$, C. Natalino$^{2}$, M. Ranjbar Zefreh$^{3}$, S. Yan$^{4}$, H. Rabbani$^{5}$, Maite Brandt-Pearce$^{5}$,J.P. Fernández-Palacios$^{6}$,\\ J.M. Rivas-Moscoso$^{6}$, O. Gonz\'{a}lez de Dios$^{6}$, J.A. Hernández$^{1}$, A. S\'anchez-Maci\'an$^{1}$, D. Larrabeiti$^{1}$, and P. Monti$^{2}$}

\IEEEauthorblockA{\textit{$^{1}$Dept. of Telematic Engineering, Universidad Carlos III de Madrid, 28911, Leganes, Madrid, Spain,}\\
$^{2}$Dept. of Electrical Engineering, Chalmers University of Technology, 412 96 Gothenburg, Sweden,\\
$^{3}$CISCO Systems S.R.L., Vimercate (MB), Italy,
$^{4}$HPN Group, Smart Internet Lab, University of Bristol, Bristol, UK,\\
$^{5}$Dept. of ECE, University of Virginia, Charlottesville, VA 22904 USA,
$^{6}$Telefónica Global CTIO, S/N, 28050 Madrid, Spain.\\
Email: \{farhad.arpanaei@uc3m.es\}}}

% \author{\IEEEauthorblockN{ M. Reza Dibaj$^{2}$, Pouya Mehdizadeh$^{2}$, M. Sadegh Ghasrizadeh$^{2}$, Hamzeh Beyranvand$^{2}$, Mahdi Ranjbar Zefreh$^{3}$,\\
% Juan Carlos Hernandez-Hernandez$^{1}$, José Alberto Hernández$^{1}$, David Larrabeiti$^{1}$, Farhad Arpanaei$^{1}$}
% \IEEEauthorblockA{\textit{$^{1}$Department of Telematic Engineering, Universidad Carlos III de Madrid (UC3M), 28911, Leganes, Madrid, Spain.}\\
% \textit{$^{2}$Department of Electrical Engineering, Amirkabir University of Technology (Tehran Polytechnic), Tehran, Iran.}\\
% \textit{$^{3}$CISCO Systems S.R.L., Vimercate (MB), Italy.} \\
% Email: \{farhad.arpanaei@uc3m.es\}}}

% \author{\IEEEauthorblockN{ M.R.Dibaj$^{1}$, P. Mehdizadeh$^{1}$, M. S. Ghasrizadeh$^{1}$, H.Beyranvand$^{1}$, D.Larrabeiti$^{2}$, F.Arpanaei$^{2}$}
% \IEEEauthorblockA{\textit{$^{1}$Department of Electrical Engineering, Amirkabir University of Technology (Tehran Polytechnic), Tehran, Iran.}\\\textit{$^{2}$Department of Telematic Engineering, Universidad Carlos III de Madrid, 28911, Leganes, Madrid, Spain.}\\
% Emails: \{farpanae@it.uc3m.es\}}}

\maketitle

\begin{abstract}
In the ultra-low inter-core crosstalk (UL-ICXT) working zone of terrestrial multi-band and multi-core fiber (MCF) elastic optical networks (EONs), the ICXT in all channels of all cores is lower than the ICXT threshold of the highest modulation format level (i.e., 64QAM) for long-haul distances (i.e., 10,000 km). This paper analyzes the performance of this type of MCF in multi-band EONs (MB-EONs). Two band and space division multiplexing (BSDM) scenarios are investigated: MCF and a bundle of multi-fiber pairs (BuMFP). Moreover, the performance of the UL-ICXT of two MCFs—one with the standard cladding diameter (CD = 125 $\mu$m) with 4 cores and another with a nonstandard larger CD with 7 cores—are evaluated in the US backbone network. Our findings show that by accurately designing the physical structure of the MCF, even with a standard CD, it is possible to achieve UL-ICXT in C+L+S-band long-haul BSDM EONs. Furthermore, the simulation results show that the network throughput for BSDM EONs with MCFs in the UL-ICXT regime is up to 12\% higher than the BuMFP scenario. Additionally, capacity increases linearly with the number of cores.%increasing the number of cores by 75\% leads to a proportional increase in capacity.   
\end{abstract}


\section{Introduction}

With the increasing demand for bandwidth, communication service providers (CSPs) are compelled to enhance their networks \cite{ciscoWhitePaper}. The adoption of multi-band technology has emerged as a cost-effective solution to address this need \cite{ArpanaeiECOC2023}. However, forecasts indicate that by 2030, the evolution of the 6th generation (6G) mobile communication ecosystem and the proliferation of artificial intelligence applications will necessitate a transition to spatial multiplexing networks \cite{WinzerParallelism2022}. Designing a flexible and versatile network that integrates band and space division multiplexing (BSDM) technologies poses a complex challenge. To meet this challenge while ensuring the stability of CSP's infrastructure and compatibility with legacy, recent proposals have introduced multi-core fibers (MCFs) with a cladding diameter identical to traditional single-core fibers, typically 125 $\mu m$ in cladding diameter \cite{MatsuiJLT2020, Matsui2015}.

However, reducing the cladding diameter (CD) decreases the distance between embedded cores, leading to increased ICXT \cite{UncoupledMCF4ECOC2020}. Effective crosstalk management requires limiting the number of cores within the standard CD. Moreover, managing crosstalk becomes critical, especially in the absence of multi-input multi-output (MIMO) digital processing along a lightpath (LP) in the uncoupled or weakly coupled MCFs \cite{Mu_OPticExpress22}. The crosstalk between adjacent cores must be managed to facilitate all-optical transmission without relying on MIMO digital processing \cite{Nakajima2017MCFStandard}. This management is influenced by factors such as transmission distance, distance between adjacent cores, number of adjacent cores, and modulation format level \cite{Mu_OPticExpress22}.

The complexity of crosstalk management is further compounded in multi-band systems due to phenomena like inter-channel stimulated Raman scattering (ISRS), which significantly impacts transmission quality based on frequency as well. Additionally, the frequency-dependent power coupling coefficient across different bands, unlike C-band systems, cannot be disregarded \cite{Matsui2015}.
Recent advancements have proposed four-core fibers of standard gauge to operate across all bands \cite{UncoupledMCF4ECOC2020}.
In this work, we not only evaluate the performance of these fibers against non-standard cladding diameter fibers ($<$ 230 $\mu m$) but also compare UL-IXCT MCFs systems with a bundle of single-mode fiber pairs (BuMFP). Furthermore, determining the suitable operating range of MCFs based on the distance between adjacent cores and trench width is crucial \cite{Ye_14, Kopp_23}. The suitable operating range ensures that crosstalk resulting from core proximity does not compromise the bit rate of transceivers, thus relying solely on linear and non-linear noise factors for determining bit rates. This working zone is called the UL-ICXT zone, where the modulation format level selection is independent of the ICXT. Therefore, network planning is not as complex as ICXT-aware service provisioning which is necessary for ICXT-sensitive weakly coupled MCFs. Moreover, the total capacity of the network is higher than that of weakly-coupled MCFs that are ICXT-sensitive. In ICXT-sensitive MCFs, the modulation format of each lightpath depends on the ICXT experienced \cite{Tetsuya_HAYASHI2014}.

% Differences in fiber parameters such as attenuation coefficient, effective area, and dispersion coefficient between multi-core and single-core fibers yield distinct outcomes.

\section{Physical Layer Modeling for band and Space division multiplexing Systems}

In this work, we consider a multi-layer optical transport network (OTN) switching-based MB-EON over MCFs or BuMFPs. We use C+L+S-band technology, providing approximately 20 THz of bandwidth. The modulation format of the line cards is adjusted based on a probabilistic constellation-shaping approach. Therefore, the modulation format of each LP depends on the channel and core used by the LP. Therefore, the bit rate of each line card can be adaptively changed based on the generalized mutual information (GMI), which is a function of the signal-to-noise ratio \cite{Bosco2019}. 

\begin{figure}[!t]
    \centering
    \includegraphics[width=1\linewidth]{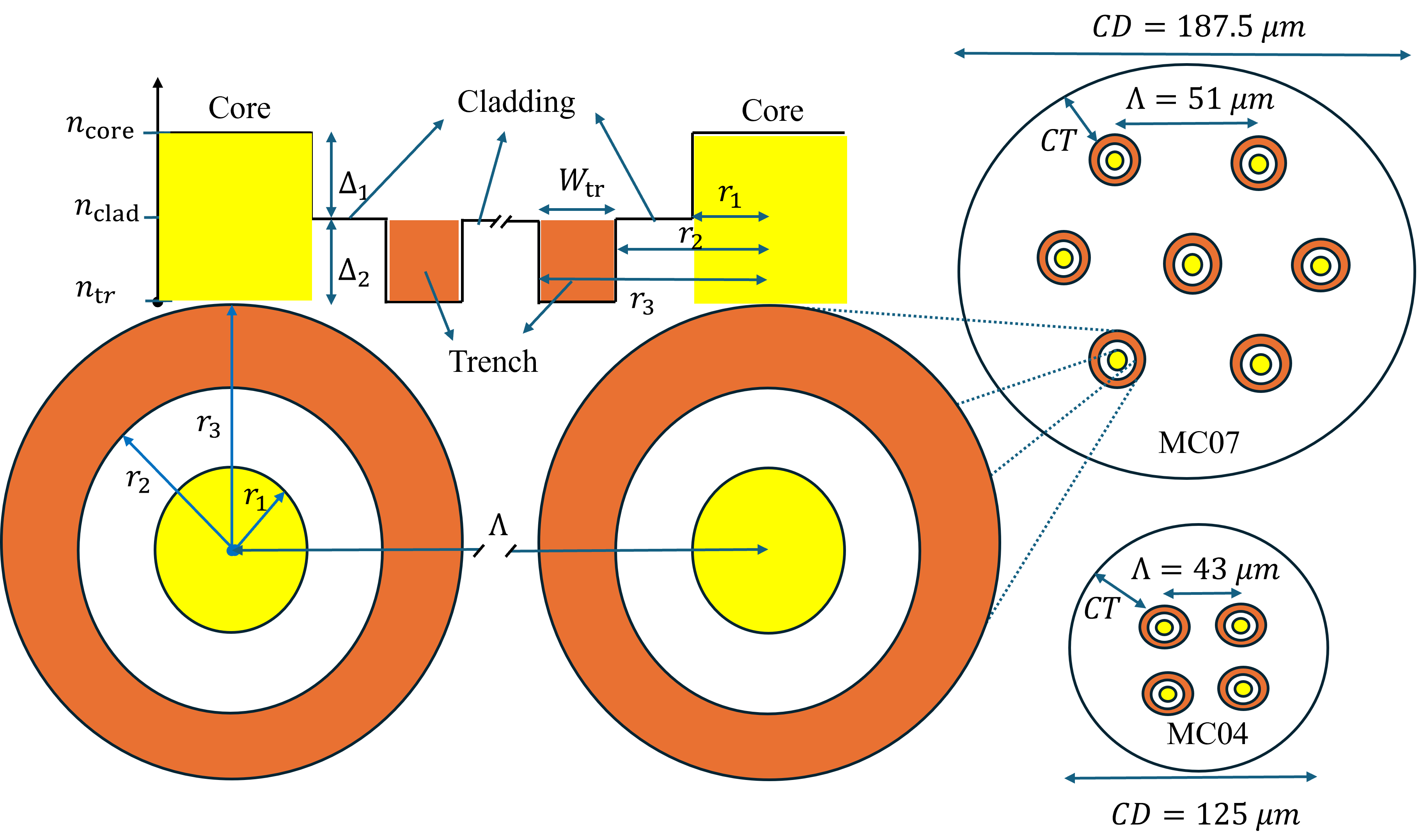}
    \caption{ Trench-assisted multi-core fiber structure, 7-core fiber (MC07) (top-right) with cladding diameter 225 $\mu m$, and 4-core fiber with cladding diameter 125 $\mu m$.}
    \label{fig_Fig1}
\end{figure}

Additionally, we assume that each link is equipped with cutting-edge technology, specifically, UL-ICXT trench-assisted (TA) MCFs. Figure \ref{fig_Fig1} illustrates the physical structure of each TA-MCF. As depicted in this figure, a trench area with a lower relative refractive index ($\Delta_2$) compared to the cladding is considered around each core, with a width of $W_{tr}$. The ICXT in TA-MCFs depends on fiber parameters such as the relative refractive index difference between the trench and cladding, trench width, and center core-to-core distance, referred to as pitch ($r_{1}$). It also depends on the frequency, which is negligible in C-band optical networks. However, as shown later, it is not negligible in multi-band systems like the C+L+S-band technology, which is the considered system model in this work.

To estimate the ICXT of TA-MCFs several works proposed numerical simulations and experimental measurements \cite{Takenaga2011}. However, authors in \cite{Ye_14} proposed an analytical closed-form model for mode coupling coefficient of the TA-MCFs. In this context, the ICXT of a TA-MCF can be calculated from \eqref{eq_mio_ICXT2}-\eqref{eq_MCC}.

 \begin{equation}\label{eq_mio_ICXT2}
	\mu_{_{\text{ICXT}}} (f^i)= \frac{N_{\text{AC}}-N_{\text{AC}}\exp[-(N_{\text{AC}}+1)\Omega(f^i) L]}{1+N_{\text{AC}}\exp[-(N_{\text{AC}}+1)\Omega(f^i) L]}, 
\end{equation}

Furthermore, the power coupling coefficient (PCC), denoted as $\Omega(f^i)$, is calculated from \eqref{eq_PCC}, where $\kappa$, $r_{\text{b}}$, $c$, $\varLambda$, $L$, $n_{\text{core}}$. $f^i$ and $N_{\text{AC}}$ represent the mode coupling coefficient (MCC), bending radius, propagation velocity, the distance between the centers of two adjacent cores (or core pitch), transmission distance, effective refractive index of the core, channel's center frequency, and the number of lit adjacent cores of the channel under test, respectively.  

\begin{equation}\label{eq_PCC}
	\Omega(f^i) = \frac{c\kappa^{2}r_{\text{b}}n_{\text{core}}}{\pi f^{i}\varLambda}
\end{equation}

Moreover, the mode coupling coefficient (MCC) is calculated from \eqref{eq_MCC}.

\begin{multline}\label{eq_MCC}
	\kappa (f^i)\cong \frac{\sqrt{\Gamma \Delta_1}}{r_1}
 \frac{U_1^2(f^i)}{V_1^3(f^i) K_1^2(W_1)} \frac{\sqrt{\pi r_1}}{W_1 \varLambda}\\ exp\{-\frac{W_1 \varLambda + 1.2[1 + V_1(f^i)] w_{\text{tr}}}{r_1}\}
\end{multline}

where the following variables are used:
\[
\Gamma = \frac{W_1}{W_1 + \frac{1.2(1 + V_1(f^i)) w_{\text{tr}}}{\varLambda}}
\]
\[
V_1(f^i) = \frac{2\pi f^i r_1 n_{\text{core}} \sqrt{2 \Delta_1}}{c}
\]
\[
K_1(W_1) = \sqrt{\frac{\pi}{2W_1}} e^{-W_1}
\]
\[
U_1^2(f^i) = \left[\frac{2 \pi f^i r_1}{n_{\text{core}} c}\right]^2 (n^4 - 1)
\]
\[
W_1|_{\frac{\Delta_2}{\Delta_1}=2} = 1.143V_1(f^i)-0.22
\]

It should be noted that in many works within the literature of the optical network planning community, the authors did not consider the dependency of the PCC and MCC on frequency, which affects the ICXT analysis in SDM-EONs \cite{Chen2023,Ma2023,Ravipudi2023, Seki2023, Zheng2022, Luo2022, ArpanaeiECOC2019, ArpanaeiJOCN2020}. They have only used the PCC at 1550 nm. For BSDM-EONs, the frequency dependency of the PCC and MCC is not negligible and must be taken into account. Therefore, the end-to-end QoT of an LP in terms of the generalized signal to the noise ratio (GSNR) at channel $i$ is estimated from \eqref{eq_GSNR_total} inspired from the incoherent Gaussian noise (GN)/enhanced GN (EGN) model for multi-band systems proposed in \cite{zefreh2020realtime}.  

\begin{multline}\label{eq_GSNR_total}
GSNR^{i}_{\text{LP}}|_{\text{dB}} = 10 \log_{10} \Big[\big(SNR_{\text{ASE}}^{-1}+SNR_{\text{NLI}}^{-1}+\\SNR_{\text{ICXT}}^{-1}+
 SNR_{\text{TRx}}^{-1}\big)^{-1} \Big]-\sigma_{\text{Flt}}|_{\text{dB}} - \sigma_{\text{Ag}}|_{\text{dB}},
\end{multline}
where $SNR_{\text{ASE}}= \Sigma_{s\in S}P_{\text{tx}}^{s+1,i}/P^{s,i}_{\text{ASE}}$, $SNR_{\text{NLI}}= \Sigma_{s\in S}P_{\text{tx}}^{s+1,i}/P^{s,i}_{\text{NLI}}$, $SNR_{\text{ICXT}}= \Sigma_{s\in S}P_{\text{tx}}^{s+1,i}/P^{s,i}_{\text{ICXT}}$ . Moreover, $P_{\text{tx}}^{s+1,i}$ is the launch power at the beginning of span $s+1$, $P^{s,i}_{\text{ASE}}= n_{\text{F}}hf^i(G^{s,i}-1)R_{\text{ch}}$ is noise power caused by the doped fiber amplifier (DFA) equipped with the digital gain equalizer (DGE), and non-linear interference (NLI) noise power ($P^{s,i}_{\text{NLI}}$) is calculated from (2) in \cite{RanjbarECOC22022}. Moreover, 
$n_{\text{F}},\,h,\,f^i,\,G^{s,i}=P_{\text{tx}}^{s+1,i}/P_{\text{rx}}^{s,i},\, S$, and $R_{\text{ch}}$ are the noise figure of DFA, Plank’s coefficient, channel frequency, frequency center of the spectrum, the gain of DFA, set of spans, and channel symbol rate, respectively. $P_{\text{rx}}^{s,i}$ is the received power at the end of span $s$. $SNR_{\text{TRx}}$, $\sigma_{\text{Flt}}$, $\sigma_{\text{Ag}}$ are the transceiver SNR, SNR penalty due to wavelength selective switches filtering, and SNR margin due to the aging. Hence, the GSNR for all potential connections from arbitrary sources to destinations in the network can be computed. Subsequently, the GMI profiles of the K shortest (path, channel, core) tuples are pre-calculated, employing GSNR thresholds for each GMI \cite{Bosco2019}.
To calculate the power of ICXTs due to the coupling of adjacent cores for MCF, we can apply $P^{s,i}_{\text{ICXT}} = \mu_{_{\text{ICXT}}}^{s,i}P_{\text{tx}}^{s+1,i}$.

% \begin{equation}\label{eq_P_ICXT}
% 	P^{s,i}_{\text{ICXT}} \cong \sum_{ c^{'} \in \mathcal{N}_{\text{AC}},c^{'}\neq c} \mu_{_{\text{ICXT}}}^{c',l,s}P_{\text{tx}}^{f,c',r,l,s}
% \end{equation}

% where $\mathcal{N}_{\text{AC}}$ is the core vicinity set of core $c$. Additionally, for a homogeneous WC-MCF in which all physical parameters of the cores are the same, we can write $P^{f,c,r,l,s}_{\text{ICXT}}= N_{\text{AC}} \mu_{_{\text{ICXT}}}^{l,s}P_{\text{tx}}^{f,c',r,l,s}$, where $N_{\text{AC}}$ is the number of lit adjacent cores of the channel under test.

A BSDM-EON over MCFs does not need the MIMO DSP equipped transceivers when the accumulated ICXT penalty on the SNR is equal or lower than 1 dB \cite{winzer2011penalties, Winzer2013, Tetsuya_HAYASHI2014}. The ICXT threshold of each GMI in terms of acceptable bit error rate (BER) according to the corresponding GSNR in dB based on \eqref{eq_snr1[dB]} is as follows.
\begin{equation}\label{eq_snr1[dB]}
	\mu_{\text{ICXT}_{\text{th}}}^m = 10\log_{10}(\frac{1-10^{(\frac{-\Gamma}{10})}}{\chi^{m}\times10^{(\frac{G_{\mathsf{th}}}{10})}}),
\end{equation}	
where $\mu_{\text{ICXT}_{\text{th}}}^m$ is the threshold of acceptable ICXT of $GMI_m$ for a given QoT penalty $\Gamma$ and BER. Additionally, $\chi^{m=1}$ to $\chi^{m=6}$ equal $0.5,\,1,\,3.41,\,5,\,10,\,21$ \cite{Tetsuya_HAYASHI2014}. 

\section{Simulations Setup, Numerical Results, and Conclusion}
\label{sec_Simulation}

\begin{figure}[!t]
    \centering
    \includegraphics[width=1\linewidth]{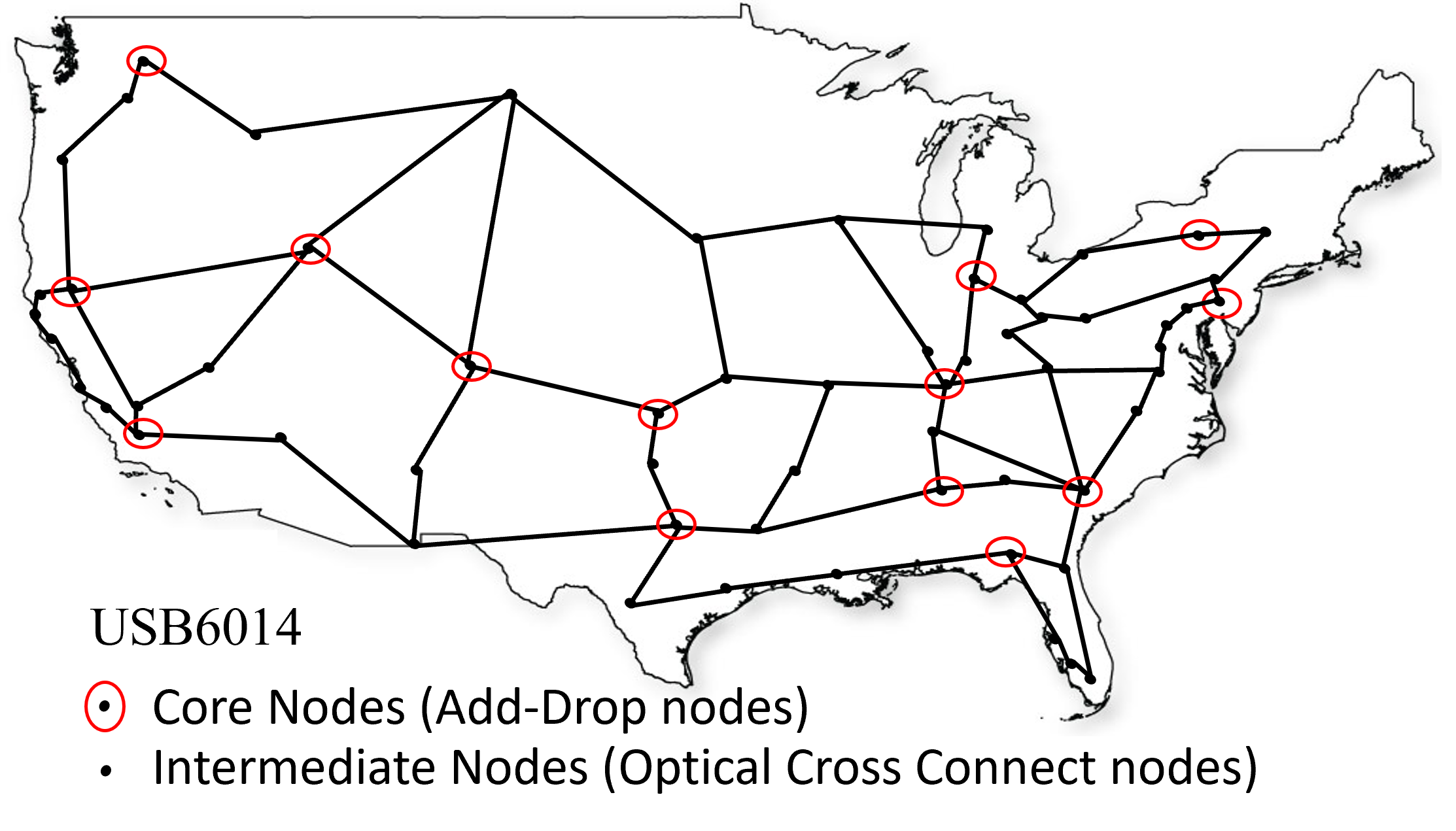}
    \caption{ Cumulative path-connection throughput for each connection in USB60.}
    \label{fig_Fig9}
\end{figure}

According to Figure \ref{fig_Fig1}, two real-world UL-ICXT TA-MCFs can be considered for the BSDM EON, namely MC07 and MC04. Additionally, standard single-mode fibers (SSMF) are used in the BuMFPs strategy. The physical optical fiber parameters are listed in Table \ref{tab_parameters}.
\begin{table}[!t]
\caption{physical parameters of the optical fibers}
\label{tab_parameters}
\begin{tabular}{|l|c|c|c|c|}
\hline
Parameter                                                                     & Symbol & MCF07 & MC04 & SSMF \\ \hline
Core counts                                                                   & $n_{\text{c}}$&      7 &   4   &   1   \\ \hline
Cladding diameter [$\mu$m]                                                            & $CD$     &  187.5     &  125    &  125    \\ \hline
Cladding thickness [$\mu$m]                                                            & $CT$ &   40    &   40   &  40    \\ \hline
Core radius  [$\mu$m]                                                                 & $r_1$  &   4.5    &   4.5   &  9    \\ \hline
Core-trench                                                                   & $r_2$  &   2$r_1$    &   2$r_1$   &  -    \\ \hline
Trench's width                                                                & $w_{\text{tr}}$&   (1,1.5,2)$r_1$    &  (1,1.5,2)$r_1$    &  -    \\ \hline
Effective area                                                                & $A_{\text{eff}}(f^i)$     &   \cite{Hayashi_12}     &   \cite{UncoupledMCF4ECOC2020}   &  \cite{RanjbarECOC22022}    \\ \hline
Loss coefficient                                                              &  $\alpha(f^i)$      &    \cite{Hayashi_12}     &    \cite{UncoupledMCF4ECOC2020}    &  \cite{RanjbarECOC22022}     \\ \hline
\begin{tabular}[c]{@{}l@{}}Effective \\ dispersion\\ coefficient\end{tabular} &  $\beta_{\text{eff}}(f^i)$      &   \cite{Hayashi_12}      &   \cite{UncoupledMCF4ECOC2020}     & \cite{RanjbarECOC22022}      \\ \hline
Core pitch  [$\mu$m]                                                                   & $\varLambda$        &   50    &   43   &  -    \\ \hline
\begin{tabular}[c]{@{}l@{}}Core's \\ refractive \\ index\end{tabular}         &   $n_{\text{core}}$     &  1.44     &  1.44    &   1.44   \\ \hline
bending radius [mm]                                                                &   $r_b$     & 144      &  144    &   140   \\ \hline
\end{tabular}
\end{table}
For the network-wide study, we consider the US backbone network with 60 nodes and 79 links, focusing on traffic exchange between the core nodes, i.e., only 14 nodes, as shown in Figure \ref{fig_Fig9}.
The average nodal degree is 2.6, the average link distance is 447 km, and the average LP distance for the shortest path (k=5) is 9534 km. In the other nodes, we have only optical cross-connects, with add/drop functionality exclusively at the core nodes. This network is called USB6014. Initially, we present a brief study on the ultra-low ICXT TA-MCFs. Subsequently, we analyze the network performance for the two SDM scenarios. 
\subsection{ TA-MCFs UL-ICXT Analysis}
\label{sec_Ultra-low}
\begin{figure*}[!t]
    \centering
    \includegraphics[width=1\linewidth]{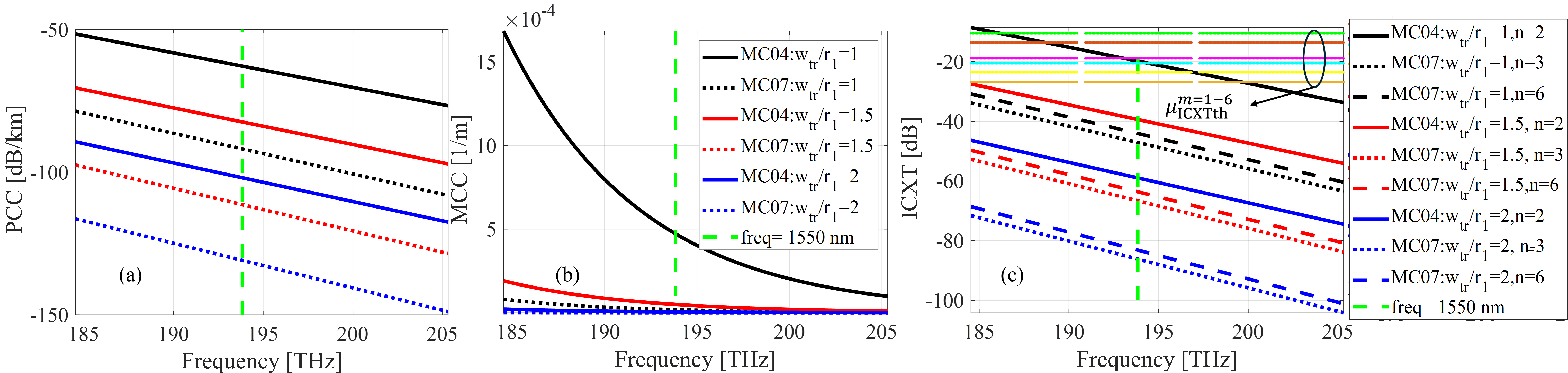}
    \caption{(a) Power coupling coefficient (PCC), (b) mode coupling coefficient (MCC), and (c) inter-core crosstalk (ICXT) versus frequency for different values of the ratio of trench width to core radius, i.e., $w_{\text{tr}}/r_1$ for seven-core MCF (MC07) and four-core MCF (MC04). For the legend of (a), please see (b).  }
    \label{fig_Fig14}
\end{figure*}

The main reason to implement TA-MCFs with ultra-low ICXT, especially in long-distance networks, is to increase the transmission bit rate and reduce planning complexity. Additionally, the ICXT in MB-EONs is frequency dependent. To demonstrate this phenomenon, let's first analyze the ICXT in terms of the PCC and MCC based on the ratio of trench width to core radius.
Figure \ref{fig_Fig14} (a) illustrates the PCC in terms of frequency in the C+L+S-band from solving \eqref{eq_PCC}. As shown in this figure, the PCC undergoes significant changes based on frequency. Therefore, in a multi-band system, we cannot consider the PCC only at 1550 nm. We can also see that the ICXT depends on the ratio of trench width to core radius, i.e., $w_{\text{tr}}/r_1$. By increasing this ratio, the ICXT decreases. However, we cannot increase it indefinitely. The distance between two adjacent trenches must be less than 3 $\mu$m \cite{Takenaga2011}. For example, for MC04, this ratio cannot exceed 2. As expected, the PCC of the MC04 is higher than the MC07 because of the higher core pitch. 

To calculate the MCC, we apply \eqref{eq_MCC}. The MCC for different values of the ratio of trench width to core radius is illustrated in Figure \ref{fig_Fig14} (b). The results show that by increasing the $w_{\text{tr}}/r_1$ from 1 to 1.5 in an MCF with a 125 $\mu$m cladding diameter, a significant decrease in MCC occurs. However, increasing it further does not result in any significant change. Regarding MC07, the results indicate that changes in $w_{\text{tr}}/r_1$ do not significantly affect the MCC. In the UL-ICXT regime, for all (path, channel, core) tuples, the ICXT must be lower than the crosstalk threshold of the highest GMI in the network, i.e., $\mu_{\text{ICXT}_{\text{th}}}^m=6 > \mu_{\text{ICXT}}$. According to \eqref{eq_snr1[dB]}, $\mu_{\text{ICXT}_{\text{th}}}^m$ is -10.58, -13.59, -18.93, -20.58, -23.59, and -26.82 dB for $m$ = 1-6, corresponding to GMI values from 2 to 12. These values are related to the pre-forward error correction (FEC) BER of \(1.5 \times 10^{-2}\). The soft decision FEC with a maximum overhead of 28\% is used \cite{Bosco2019}.  The ICXT based on \eqref{eq_mio_ICXT2} is simulated for a transmission reach of 10,000 km using the parameters in Table \ref{tab_parameters}. As depicted in Figure \ref{fig_Fig14} (c), we can conclude several findings. Firstly, the ICXT of a channel depends not only on the transmission reach but also on the frequency, physical structure of the MCF, number of adjacent cores, and core pitch. For example, in the MC04, as represented in Figure \ref{fig_Fig1}, each core has two adjacent cores and a core pitch of 43 \(\mu\)m. Despite this, the ICXT in the MC07 cores is lower than in the MC04. Although the MC07 has two types of cores—the inner ones with 6 neighbors and the outer ones with 3 neighbors—the dominant effect on the ICXT is the core pitch, which is 51 \(\mu\)m in the MC07. Therefore, if we consider an MC07 with a smaller CD, we have to decrease the pitch, resulting in more ICXT. 

It should be noted that because of fabrication restrictions, an MC07 with a standard CD of 125 \(\mu\)m and a core pitch of 43 \(\mu\)m is not possible. Furthermore, increasing the CD increases the effective area; for example, in the C-band, it is about 80 \(\mu\)m\(^2\) and 120 \(\mu\)m\(^2\) for MC04 and MC07, respectively. The primary challenge in designing MCFs is optimizing the number of cores within a limited cladding. Currently, there is no standard cladding size for MCFs, but a smaller cladding diameter is preferred to achieve high core density and ensure strong mechanical reliability when bending. To meet the failure probability limits, the cladding diameter should be less than 230 \(\mu\)m. Considering this fabrication restriction, the MC07 with a core pitch of 51 \(\mu\)m with $CD=187.5 \mu $m is an optimum choice for MCFs with a hexagonal close-packed structure \cite{Hayashi_12}. 

Additionally, based on Figure \ref{fig_Fig14} (c), the ICXT for all channels and cores is lower than the ICXT thresholds of the modulation format level with the highest GMI, even after a 10,000 km transmission reach. However, we do not have this situation for the MC04 with \(w_{\text{tr}}/r_1=1\). For these types of MCFs, like the MC04 with \(w_{\text{tr}}/r_1=1\), a complicated ICXT-aware algorithm for service provisioning and network planning is required. By increasing the \(w_{\text{tr}}/r_1\) to 1.5, we achieve a good working zone for the MC04, where the ICXT for all channels in all cores is lower than -26.82 dB. Therefore, to have a fair comparison between MC04 and MC07, we consider \(w_{\text{tr}}/r_1=1.5\).
% \begin{figure} [!t]
%     \centering
%     \includegraphics[width=1\linewidth]{Fig12}
%     \caption{Inter-core crosstalk (ICXT) v.s frequency for different values of the ratio of trench width to core radius, i.e., $w_{\text{tr}}/r_1$ for seven-core MCF (MC07) and four-core MCF (MC04).}
%     \label{fig_Fig12}
% \end{figure}
\subsection{ Long-haul networks over ultra-low ICXT TA-MCFs and BuMFPs }
\label{sec_Net}
\begin{figure}[!t]
    \centering
    \includegraphics[width=1\linewidth]{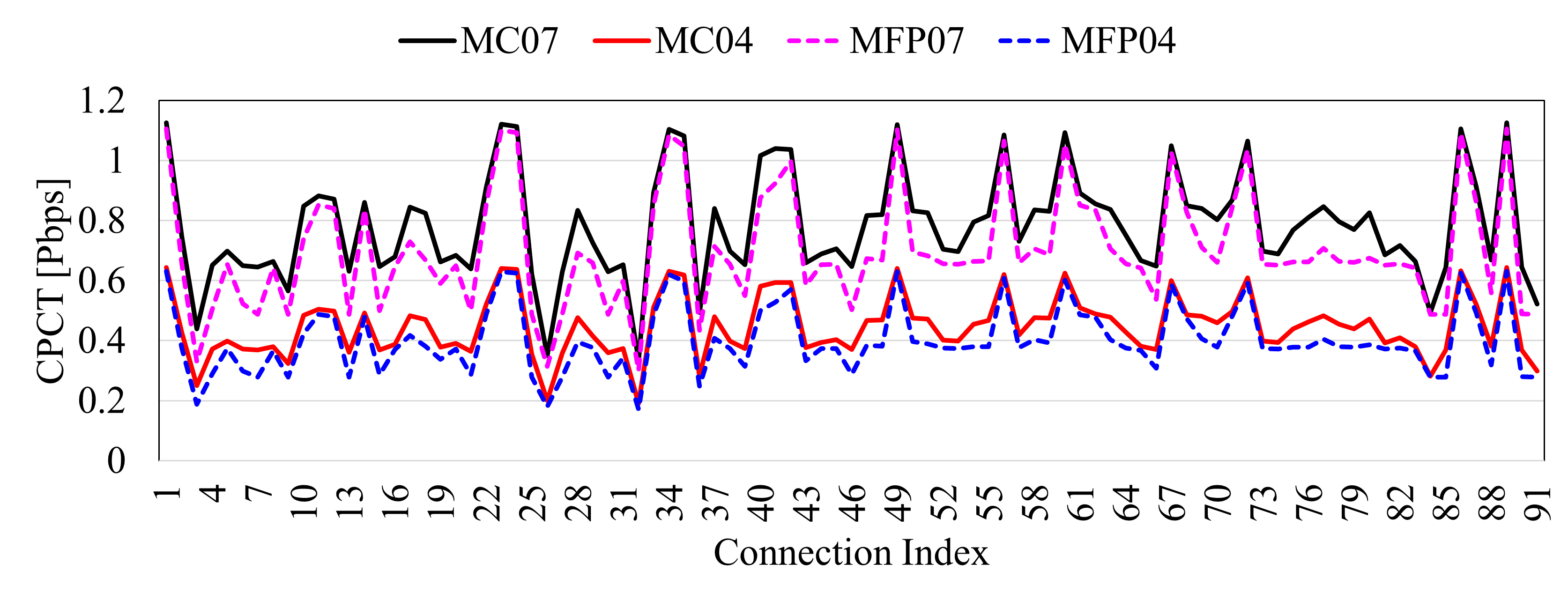}
    \caption{ Cumulative path-channel throughput (CPCT) for each connection in USB60.}
    \label{fig_Fig8}
\end{figure}

\begin{figure}[!t]
    \centering
    \includegraphics[width=1\linewidth]{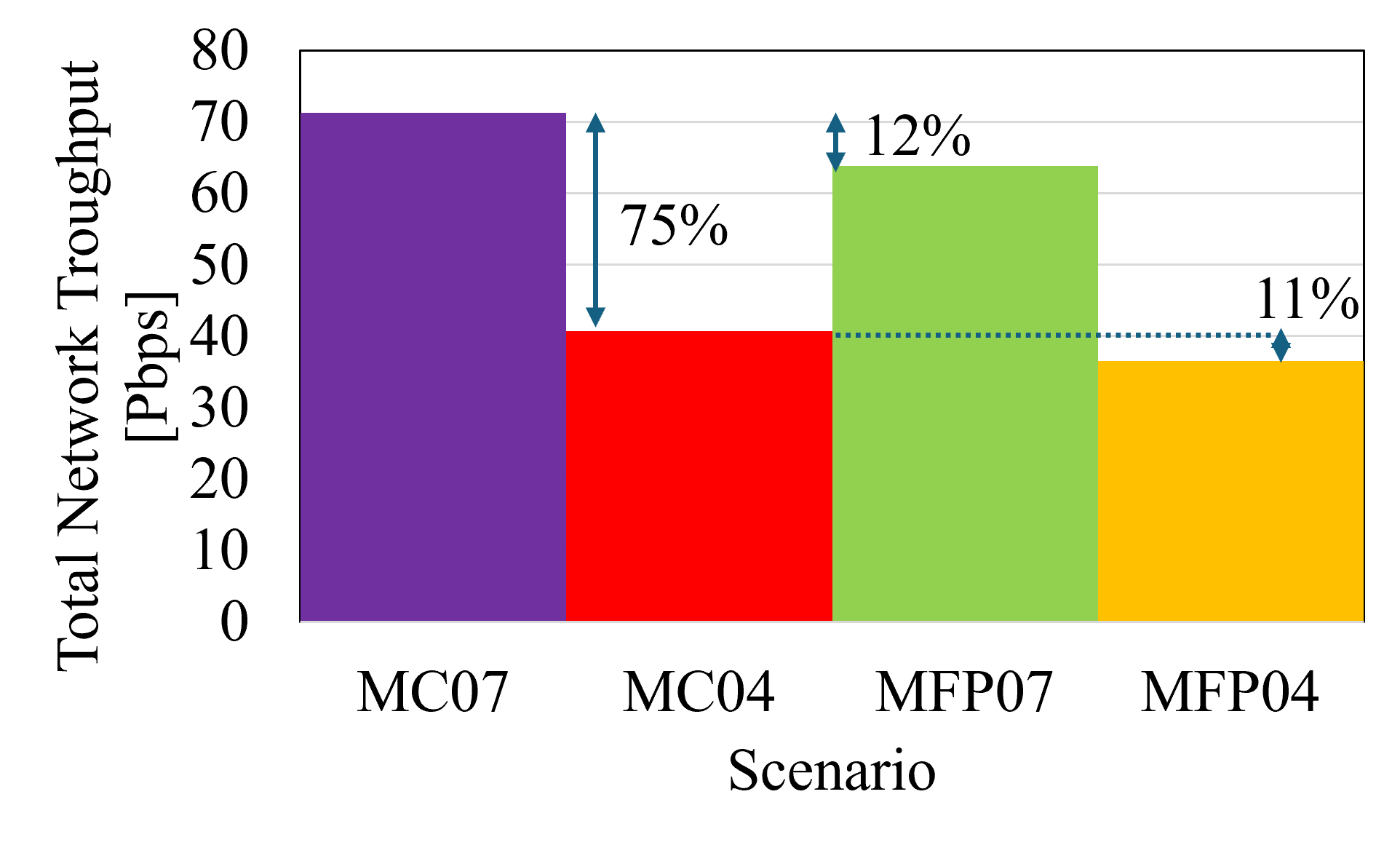}
    \caption{ Cumulative path-connection throughput for each connection in USB60.}
    \label{fig_Fig13}
\end{figure}

In this section, we investigate two BSDM strategies, namely MCFs and BuMFPs. However, inspired by the previous discussion, we focus on UL-ICXT MCFs. All (path, channel, core) tuples' transmission bit rates for the USB6014 over a C+L+S-band with 268 channels of 75 GHz bandwidth are calculated for 92 connections in the network. The maximum span length is 80 km in each link. The symbol rate of each channel is 64 GBaud, and the transmission bit rate varies between 100 Gbps and 600 Gbps based on the GSNR of each channel, calculated according to \eqref{eq_GSNR_total} for the k=1 shortest path of each connection. The noise figures of the DFA amplifiers are 4.5 dB, 5 dB, and 6 dB in the C-, L-, and S-band, respectively. The optimum pre-tilted launch power for each span is calculated based on the hyper-accelerated scheme introduced in \cite{ArpanaeiHPO_OFC2024}.  
The cumulative throughput of all (path, channel, core) tuples for each connection is illustrated in Figure \ref{fig_Fig8}. The results reveal that the performance disparity between MFC and BuMFP. The main reason for this is that the loss coefficient of the MCFs is lower than that of the SSMF, as shown in Table \ref{tab_parameters}. To gain a more insightful understanding, we calculate the total network throughput as illustrated in Figure \ref{fig_Fig13}. As shown in Figure \ref{fig_Fig13}, not only do the UL-IXCT MCFs have higher throughput (about 12\%) compared to BuMFP, but the increased rate of throughput is proportional to the spatial lanes (cores/fiber pair) in both scenarios. For example, by increasing the core/fiber counts by 75\%, the throughput also increases by 75\%. A techno-economic study based on the pay-as-you-grow approach is suggested for the next study.

\bibliographystyle{IEEEtran}  % You can change this to the style you prefer
\bibliography{references}

\end{document}